\documentclass[pre,onecolumn,floatfix,showpacs]{revtex4}
\usepackage{graphicx,bm,amsmath}
\begin{document}
\title{Effects of an oscillating field on pattern formation in a
ferromagnetic thin film: Analysis of patterns traveling at a low velocity}

\author{Kazue Kudo}
\email{kudo@a-phys.eng.osaka-cu.ac.jp}
\author{Katsuhiro Nakamura}
\affiliation{Department of Applied Physics, Graduate School of
Engineering, Osaka City University, Osaka 558-8585, Japan}

\date{\today}

\begin{abstract}
Magnetic domain patterns under an oscillating field are studied
 theoretically by using a simple Ising-like model.
We propose two ways to investigate the effects of the oscillating
 field. The first one leads to a model in which rapidly oscillating terms are
 averaged out and the model can explain the existence of the maximum
 amplitude of the field for the appearance of
 patterns. The second one leads to a model that includes the delay of the
 response to the field and the model suggests the existence of a
 traveling pattern which moves very slowly compared with the time scale
 of the driving field.
\end{abstract}


\maketitle

\section{\label{sec:intro} Introduction}

Rapidly driven systems have received  considerable attention these
days. Under a rapidly 
oscillating field, a state which is unstable in the absence of the
oscillating field can be stabilized. One of the most simple 
and well-known examples is Kapitza's inverted pendulum
and Landau and Lifshitz generalized the problem~\cite{landau}. 
Their method has recently been applied to classical and quantum dynamics in
periodically driven systems~\cite{rahav03,rahav05}.
The method is also applied to stabilization of a matter-wave soliton in
two-dimensional Bose-Einstein condensates
without an external trap~\cite{saito,abdullaev,liu}. 

Magnetic domain patterns in a uniaxial ferromagnetic thin film, which
usually show a labyrinth structure, exhibit various kinds of structures
under an oscillating field. For example, the labyrinth structure changes
into a parallel-stripe structure for a certain
field~\cite{miura,mino}. In some other cases, several types of lattice
structures can appear~\cite{tsuka}.

\begin{figure}
\includegraphics[width=8cm]{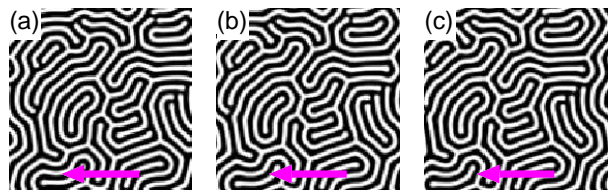}
\caption{\label{fig:trvl} Snapshots of a traveling pattern under an
 oscillating field after (a) 5000 cycles, (b) 10000 cycles, and (c)
 15000
 cycles. The whole pattern is traveling to the left. 
The details are described in Appendix~\ref{sec:app}.}
\end{figure}

In this paper, we develop effective theories for slow motion of magnetic
domain patterns under a rapidly oscillating field.
Especially, we focus on traveling patterns as an example of
slowly moving patterns. 
So far, there were few effective theories to describe such a slowly
traveling pattern under a rapidly oscillating field.
In experiments on a garnet thin film, 
we can observe a parallel-stripe pattern
traveling very slowly compared with the time scale of the field
in some cases~\cite{mino_p}. 
A traveling mazelike pattern like Fig.~\ref{fig:trvl} is also found in our 
numerical simulations. 

Although traveling patterns appear in various kinds of systems,  
most works about them have been limited to the systems
in the absence of an oscillating 
field~\cite{malomed,coullet,douady,fauvePRL,price,lega,okuzono,sagues}.
The mechanism of such traveling patterns in one-dimensional (1D) systems
was intensively studied  in the 1980s
as  drift instabilities or parity-breaking
instabilities~\cite{malomed,coullet,douady,fauvePRL}.
In Ref.~\cite{malomed}, secondary instabilities were discussed for
several similar equations. By contrast, the authors of
Refs.~\cite{coullet,douady,fauvePRL} gave no particular equation at
first, but they considered 
symmetries of the system and assumed the form of the solution before
deriving their equations.
Almost 10 years after those papers, 
Price studied traveling patterns in 2D scalar
nonlinear neural fields where the nonlinearity is purely cubic and
discussed constraints on the neural field structure and parameters to
support traveling patterns~\cite{price}. He suggested that 
Swift-Hohenberg-type models will not support traveling patterns. However,
moving patterns were actually observed in numerical simulations for a
general complex Swift-Hohenberg equation in Ref.~\cite{lega}. 
Our model has similar properties to those of the (real) Swift-Hohenberg
equation. We will not use the method of Ref.~\cite{price} but that of
Ref.~\cite{malomed} to explore the existence of a traveling pattern in
a ferromagnetic thin film.

In fact, recently, domain walls under a rapidly oscillating field have 
been studied~\cite{michaelis,kirakosyan}. Michaelis \textit{et al.} 
discussed the effects of rapid periodic oscillation of parameters in a
Ginzburg-Landau (GL) equation by applying a multiscale technique and 
derived the averaged GL equation~\cite{michaelis}; 
Kirakosyan \textit{et al.} derived the averaged Landau-Lifshitz equation 
by employing the multi-time-scale expansion technique~\cite{kirakosyan}.
In those papers, they took into account higher harmonic oscillations.
Although their methods cannot be directly applied to our model, our 
methods correspond to the lowest orders of their multi-time-scale 
expansions.

Our model is a simple 2D Ising-like model (see
Refs.~\cite{jagla04,jagla05,kudoE,kudoS}, and references therein), which
has been used to simulate magnetic domain patterns. The
numerical results simulated by the model 
show very similar properties to experimental 
ones~\cite{jagla04,jagla05,kudoE,kudoS}.
We consider a scalar field $\phi(\bm{r})$, where $\bm{r}=(x,y)$. The
positive and negative values of $\phi(\bm{r})$ correspond to the up and
down spins, respectively. The Hamiltonian consists of four energy
terms: Uniaxial-anisotropy energy $H_{\rm ani}$, exchange interactions
$H_J$, dipolar interactions $H_{\rm di}$, and interactions with the
external field $H_{\rm ex}$. The anisotropy energy is given by
\begin{equation}
 H_{\rm ani}=\alpha \int {\rm d}\bm{r} \left(
-\frac{\phi(\bm{r})^2}{2}+\frac{\phi(\bm{r})^4}{4}
\right).
\label{eq:Ha}
\end{equation}
It implies that the anisotropy energy prefers the values 
$\phi(\bm{r})=\pm 1$.
The exchange and dipolar interactions are described by
\begin{equation}
 H_J=\beta\int {\rm d}\bm{r} \frac{|\nabla\phi(\bm{r})|^2}{2}
\label{eq:Hj}
\end{equation}
and
\begin{equation}
 H_{\rm di}=\gamma\int {\rm d}\bm{r} {\rm d}\bm{r}'
 \phi(\bm{r})\phi(\bm{r}') G(\bm{r},\bm{r}'),
\label{eq:Hdi}
\end{equation}
respectively. Here, $G(\bm{r},\bm{r}')\sim |\bm{r}-\bm{r}'|^{-3}$ at
long distances.
The exchange interactions imply that $\phi(\bm{r})$ tends to have the 
same value as neighbors. On the other hand, the dipolar interactions 
imply that $\phi(\bm{r})$ tends to have the opposite sign to the values 
in a region at a long distance. 
Namely, $H_J$ and $H_{\rm di}$ may be interpreted as short-range 
attractive and
long-range repulsive interactions, respectively. Their competition
leads to a domain structure with a characteristic length.
The term from the interactions with the external field is given by
\begin{equation}
 H_{\rm ex}=-h(t) \int {\rm d}\bm{r} \phi(\bm{r}).
\label{eq:Hex}
\end{equation}
Here, we consider a spatially homogeneous and  rapidly oscillating field:
\begin{equation}
 h(t)=h_0\sin\omega t.
\label{eq:h}
\end{equation}
From Eqs.~(\ref{eq:Ha})--(\ref{eq:Hex}), the dynamical equation of the
model is described by 
\begin{eqnarray}
 \frac{\partial \phi (\bm{r})}{\partial t}&=&
 -L_0\frac{\delta (H_{\rm ani}+H_{J}+H_{\rm di}+{H_{\rm ex}})}
 {\delta \phi (\bm{r})} \nonumber\\
&=& L_0 \biggl\{ \alpha  [\phi(\bm{r})-\phi(\bm{r})^3]
+\beta\nabla^2\phi(\bm{r}) 
-\gamma\int {\rm d}\bm{r}' \phi(\bm{r}') G(\bm{r},\bm{r}')
+h(t)\bigg\}.
\label{eq:A-C}
\end{eqnarray}
Hereafter, we fix $L_0=1$ and give the parameters $\alpha$, $\beta$,
$\gamma$, and $h_0$ as positive values.

In this paper, we propose two approximation methods to obtain the dynamical
equation for slow motion. In both methods, we apply a part of Kapitza's
idea that 
the dynamics under a rapidly oscillating field can be separated into a
rapidly oscillating part and a slowly varying part~\cite{landau}. In
Sec.~\ref{sec:time}, we derive the model whose rapidly oscillating part
is averaged out on the basis of the Kapitza's idea about the time
average of the fast motion. 
In Sec.~\ref{sec:phase}, we derive another model for the slow motion,
considering the delay of the
response to the field instead of taking a time average. 
After the derivation of the models, the instabilities of
traveling patterns are investigated in both Secs.~\ref{sec:time} and
\ref{sec:phase}. We discuss the details about the existence of a
traveling pattern in Sec.~\ref{sec:disc}.
Conclusions are given in Sec.~\ref{sec:conc}.

\section{\label{sec:time} Time-averaged model}

First of all, we assume that the variable $\phi(\bm{r})$ can be
separated into two parts: 
\begin{equation}
 \phi(\bm{r},t)=\Phi(\bm{r},t)+\phi_0(t).
\label{eq:1_0}
\end{equation}
Here, $\Phi(\bm{r},t)$ is a slowly varying term and $\phi_0(t)$ is a
rapidly oscillating space-independent term.  
Substituting Eq.~(\ref{eq:1_0}) into Eq.~(\ref{eq:A-C}), we obtain
\begin{equation}
 \frac{\partial\Phi(\bm{r})}{\partial t}+\dot{\phi_0}
=\alpha \left[ (\Phi(\bm{r})+\phi_0)-
(\Phi(\bm{r})+\phi_0)^3\right] +\beta\nabla^2(\Phi(\bm{r})+\phi_0)
-\gamma\int {\rm d}\bm{r}'(\Phi(\bm{r}')+\phi_0)G(\bm{r},\bm{r}')+h(t).
\label{eq:1_1}
\end{equation}

Let us consider only the rapidly oscillating space-independent part;
then we have
\begin{equation}
 \dot{\phi_0}=\alpha (\phi_0-\phi_0^3)-
\gamma\phi_0\int {\rm d}\bm{r}' G(\bm{r}',0)+h(t).
\label{eq:1_2}
\end{equation}
Here, we define $G(\bm{r},0)\equiv 1/|\bm{r}|^3$. Then, the integral in
Eq.~(\ref{eq:1_2}) is a constant, $a_0$:
\begin{equation}
 a_0=2\pi\int_d^{\infty} \frac{{\rm d}r}{r^2}.
\label{eq:a0}
\end{equation}
Here, $d$ is the cutoff length to prevent the divergence for $d\to 0$.
It is also interpreted as the lower limit of the dipolar interactions.
The solution of Eq.~(\ref{eq:1_2}) should have the following form:
\begin{equation}
 \phi_0=\rho_0\sin (\omega t +\delta),
\label{eq:form}
\end{equation}
where $\delta$ is a phase shift which comes from the delay of the
response to the field. Substituting Eq.~(\ref{eq:form}) into
Eq.~(\ref{eq:1_2}) and omitting high-order harmonics 
(i.e. $\sin 3\omega t$), we have
\begin{equation}
 \omega\rho_0\cos(\omega t+\delta)=\eta_0\rho_0\sin(\omega t+\delta) 
-\frac34\alpha\rho_0^3\sin(\omega t+\delta) +h_0\sin\omega t,
\label{eq:1_3}
\end{equation}
where $\eta_0=\alpha-\gamma a_0$. From Eq.~(\ref{eq:1_3}), a pair of
simultaneous equations is obtained:
\begin{subequations}
\label{eq:1_4}
\begin{eqnarray}
 -\omega\rho_0\sin\delta &=& \left( \eta_0-\frac34\alpha\rho_0^2 \right)
 \rho_0\cos\delta + h_0, \\
 \omega\rho_0\cos\delta &=& \left( \eta_0-\frac34\alpha\rho_0^2 \right)
 \rho_0\sin\delta.
\end{eqnarray} 
\end{subequations}
Eliminating $\delta$ from Eq.~(\ref{eq:1_4}), we obtain a cubic equation
of $X\equiv\rho_0^2$:
\begin{equation}
 \frac{9}{16}\alpha^2 X^3-\frac32\alpha\eta_0 X^2 +(\omega^2+\eta_0^2)X
=h_0^2.
\label{eq:1_5}
\end{equation}
Therefore, $\rho_0$ can be evaluated from Eq.~(\ref{eq:1_5})
if the parameters $\alpha$, $\eta_0$, $\omega$, and $h_0$ are given.

Now, let us think about the slowly varying part. After substituting
Eq.~(\ref{eq:form}) into Eq.~(\ref{eq:1_1}), we average
out the rapid oscillation, closely following Kapitza's idea. 
Then, we obtain an equation for slowly
varying domain patterns:
\begin{equation}
 \frac{\partial\Phi(\bm{r})}{\partial t}
=\alpha \left( \Phi(\bm{r}) - \Phi(\bm{r})^3 \right)
  -\frac32\alpha\rho_0^2\Phi(\bm{r}) 
+\beta\nabla^2\Phi(\bm{r}) -\gamma\int {\rm d}
\bm{r}'\Phi(\bm{r}')G(\bm{r},\bm{r}'). 
\label{eq:t_av}
\end{equation}
The second term on the right hand side of Eq.~(\ref{eq:t_av}) is 
an extra term due to the time average.
This term is essential to explore the effects of the rapidly oscillating
field. 

On the basis of Eq.~(\ref{eq:t_av}), we will analyze the possibility of
patterns traveling at a low velocity.
Let us first choose the most simple moving stripe-type solution for 
Eq.~(\ref{eq:t_av}):
\begin{equation}
 \Phi(\bm{r},t)=A_0(t)+A_1(t)\sin(kx+b(t)).
\label{eq:2_0}
\end{equation} 
Substituting Eq.~(\ref{eq:2_0}) into Eq.~(\ref{eq:t_av}) and omitting
high-order harmonics, we have
\begin{eqnarray}
 \dot{A}_0+\dot{A}_1\sin(kx+b)&&+\dot{b}A_1\cos(kx+b)
=\eta_0'A_0+\eta_1'A_1\sin(kx+b) \nonumber\\
&&-\alpha\left[
A_0^3+3A_0^2A_1\sin(kx+b)+\frac32A_0A_1^2
+\frac34A_1^3\sin(kx+b) \right].
\label{eq:2_1}
\end{eqnarray}
Here,
\begin{subequations}
\label{eq:2_eta} 
\begin{eqnarray}
 \eta_0'&=&\left( 1-\frac32\rho_0^2 \right)\alpha -\gamma a_0,
\\
 \eta_1'&=&\left( 1-\frac32\rho_0^2 \right)\alpha -\beta k^2 
-\gamma (a_0-a_1k),
\end{eqnarray}
\end{subequations}
with $a_0$ given by Eq.~(\ref{eq:a0}), $a_1=2\pi$, and $k=|\bm{k}|$.
Equation~(\ref{eq:2_1}) leads to the following equations:
\begin{subequations}
\label{eq:2_2}
\begin{eqnarray}
 \dot{A}_0&=&\eta_0'A_0-\alpha\left( A_0^3+\frac32 A_0A_1^2 \right),\\
 \dot{A}_1&=&\eta_1'A_1-\alpha\left( 3A_0^2A_1+\frac34 A_1^3 \right),\\
 \dot{b}&=&0. \label{eq:2_b1}
\end{eqnarray}
\end{subequations}
Equation~(\ref{eq:2_b1}) implies that the phase $b(t)$ in
Eq.~(\ref{eq:2_0}) shows no time dependence and that there is no
traveling pattern with the simplest form like Eq.~(\ref{eq:2_0}).

Next, let us consider a more generalized solution by incorporating the
second harmonics:
\begin{equation}
 \Phi(\bm{r},t)=A_0(t)+A_1(t)\sin(kx+b(t))
 +A_{21}\cos[2(kx+b(t))]+A_{22}\sin[2(kx+b(t))].
\label{eq:2_3}
\end{equation} 
Substituting Eq.~(\ref{eq:2_3}) into Eq.~(\ref{eq:t_av}) leads to the
following equations:
\begin{subequations}
 \label{eq:2_4}
\begin{eqnarray}
 \dot{A}_0&=&\eta_0'A_0-\alpha \left( A_0^3 +\frac32 A_0A_1^2 +\frac32 
  A_0A_{21}^2 +\frac32 A_0A_{22}^2 -\frac34A_1^2A_{21} \right),\\
 \dot{A}_1&=&\eta_1'A_1-\alpha \left( \frac34 A_1^3 +3A_0^2A_1 +\frac32
  A_1A_{21}^2 +\frac32 A_1A_{22}^2 -3A_0A_1A_{21} \right),\\
 \dot{A}_{21}&=&\eta_2'A_{21}-\alpha \left( \frac34 A_{21}^3 
 -\frac32 A_0A_1^2 +3A_0^2A_{21} +\frac32 A_1^2A_{21} 
 +\frac34 A_{21}A_{22}^2 -6A_0A_{22}^2 \right),\\
 \dot{A}_{22}&=&\eta_2'A_{22}-\alpha \left( \frac34 A_{22}^3 
 +3A_0^2A_{22} +\frac34 A_{21}^2A_{22} +\frac32 A_1^2A_{22} 
 +6A_0A_{21}A_{22} \right),
\end{eqnarray}
\end{subequations}
and
\begin{equation}
 \dot{b}=-3\alpha A_0A_{22}. 
\label{eq:2_b2} 
\end{equation}
Here, $\eta_0'$ and $\eta_1'$ are given by Eq.~(\ref{eq:2_eta}), and 
\begin{equation}
 \eta_2'=\left( 1-\frac32\rho_0^2 \right)\alpha 
 -4\beta k^2 -\gamma (a_0-2a_1k). 
\end{equation}
This time, Eq~(\ref{eq:2_b2}) implies that there can be a traveling
pattern if $A_0\ne 0$ and $A_{22}\ne 0$.

Now let us find a stationary point (SP) of Eq.~(\ref{eq:2_4}) 
where $A_0=0$ or $A_{22}=0$, and 
examine its linear stability. If the SP is
unstable and both $A_0$ and $A_{22}$ grow from zero, 
the pattern can start to travel.
For the parameter values used to obtain Fig.~\ref{fig:trvl},
however, there are no SPs except for ones with $A_0=A_{21}=A_{22}=0$.
We should note $A_1=0$ or $A_1^2=4\eta_1'/3\alpha$ at the
SPs with $A_0=A_{21}=A_{22}=0$.
Since $A_1$ must be real, $\eta_1'>0$. Namely,
\begin{equation}
 \rho_0^2 < \frac{2}{3\alpha}\left[
\alpha -\beta k^2 -\gamma (a_0-a_1k) \right].
\label{eq:2_5}
\end{equation}
This condition gives an estimate of the maximum value of the field
amplitude $h_0$ to observe a nonuniform pattern, as $h_0$ proves to be
a monotonic function of 
$\rho_0$ for the parameter values in Fig.~\ref{fig:trvl}. 
In other words, if Eq.~(\ref{eq:2_5}) is not satisfied, the only SP is 
$(A_0,A_1,A_{21},A_{22})=(0,0,0,0)$, which means that no pattern appears.
At the SPs with $A_0=A_{21}=A_{22}=0$ and
$A_1^2=4\eta_1'/3\alpha$, the Jacobian of Eq.~(\ref{eq:2_4})
becomes 
\begin{equation}
 J=\left(
\begin{array}{cccc}
 \eta_0'-2\eta_1' & 0 & \eta_1' & 0\\
 0 & -2\eta_1'& 0 & 0 \\
 2\eta_1' & 0 & \eta_2'-2\eta_1' & 0\\
 0 & 0 & 0 & \eta_2'-2\eta_1'
\end{array}
\right).
\label{eq:Jacobi2}
\end{equation}
The real parts of eigenvalues of Eq.~(\ref{eq:Jacobi2}) are 
$\Lambda_1=-2\eta_1'$,   
$\Lambda_2=\Lambda_3=\frac12 (\eta_0'-4\eta_1'+\eta_2')$, and 
$\Lambda_4=\eta_2'-2\eta_1'$. Note that $\Lambda_1$ is
always negative. 
The others ($\Lambda_2$, $\Lambda_3$, $\Lambda_4$)  
also prove to be negative when $k\simeq 1$.
In fact, the most preferable wave number of domain patterns is $k=1$ for
the parameter values in Fig.~\ref{fig:trvl} (see Ref.~\cite{kudoS} for
details). Therefore, the present SPs
are stable and we cannot expect a traveling pattern in this case.

\section{\label{sec:phase} Phase-shifted model}

In this section, we consider another equation for slowly varying domain
patterns instead of Eq.~(\ref{eq:t_av}). We begin with
Eqs.~(\ref{eq:1_0})--(\ref{eq:1_5}) again, but we will not take a 
time average. 
Instead, we take the delay of the response to the field into consideration.
Substituting
Eq.~(\ref{eq:form}) into Eq.~(\ref{eq:1_1}), we consider the equation as
a discrete-time
equation which is valid at $t=(2\pi/\omega)n$ with integers $n$. 
Then, we regard the discrete
time as continuous. This procedure is justified  when the field
oscillation is rapid enough compared 
with the time scale of the slowly varying part.
It is as if we take a sequence of snapshots at
$t=(2\pi/\omega)n$ and take it as a movie. 
In fact, our numerical results in Fig.~\ref{fig:trvl} are obtained by
taking these kinds of snapshots.
We thus obtain a new equation for slowly varying domain patterns:
\begin{equation}
 \frac{\partial\Phi(\bm{r})}{\partial t}=
\alpha (1-3\rho_0^2\sin^2\delta )\Phi(\bm{r}) +\beta\nabla^2\Phi(\bm{r})  
-\gamma\int{\rm d}\bm{r}'\Phi(\bm{r}')G(\bm{r},\bm{r}')
-\alpha\Phi(\bm{r})^2\left( \Phi(\bm{r})+3\rho_0\sin\delta \right) +C,
\label{eq:p_sf}
\end{equation}
where 
\begin{equation}
 C= \eta_0\rho_0\sin\delta -\alpha\rho_0^3\sin^3\delta
  -\omega\rho_0\cos\delta, 
\end{equation}
with $\rho_0$ and $\delta$ evaluated from Eq.~(\ref{eq:1_4}). 
Equation~(\ref{eq:p_sf}) has two extra terms due to the phase shift
$\delta$ except for the constant $C$. 
One is linear and the other is nonlinear in $\Phi$. 
The extra nonlinear term has an important role in discussion of 
the existence of a traveling pattern.

Now, let us consider the stability of a traveling pattern
on the basis of Eq.~(\ref{eq:p_sf}).
When the simplest form, Eq.~(\ref{eq:2_0}), 
is substituted into Eq.~(\ref{eq:p_sf}), we
obtain the same result as Eq.~(\ref{eq:2_b1}).
Therefore, we proceed to choose the extended solution, Eq.~(\ref{eq:2_3}).
Substituting Eq.~(\ref{eq:2_3}) into Eq.~(\ref{eq:p_sf}) leads to the
following equations:
\begin{subequations}
\label{eq:3_0}
\begin{eqnarray}
 \dot{A}_0&=&\tilde{\eta}_0A_0+C \nonumber\\ 
&& -\alpha\left[
A_0^2(A_0+3\rho_0\sin\delta) +\frac32 (A_0+\rho_0\sin\delta)
(A_1^2+A_{21}^2+A_{22}^2) -\frac34 A_1^2A_{21} 
\right],\\
 \dot{A}_1&=&\tilde{\eta}_1A_1 \nonumber\\
&& -\alpha\left[
\frac34 A_1^3 +\frac32 A_1(A_{21}^2+A_{22}^2)
+3A_0A_1(A_0+2\rho_0\sin\delta) -3A_1A_{21}(A_0+\rho_0\sin\delta)
\right],\\
 \dot{A}_{21}&=&\tilde{\eta}_2A_{21} \nonumber\\
&& -\alpha\left[
\frac34 A_{21}^3 +\frac34 A_{21}(2A_1^2+A_{22}^2) 
+3A_0A_{21}(A_0+2\rho_0\sin\delta) 
-\frac32 (A_1^2+4A_{22}^2)(A_0+\rho_0\sin\delta) 
\right],\\
 \dot{A}_{22}&=&\tilde{\eta}_2A_{22} \nonumber\\
&& -\alpha\left[
\frac34 A_{22}^3 +\frac34 A_{22}(2A_1^2+A_{21}^2) 
+3A_0A_{22}(A_0+2\rho_0\sin\delta) 
+6A_{21}A_{22}(A_0+\rho_0\sin\delta) 
\right],
\end{eqnarray} 
and
\begin{equation}
 \dot{b}=-3\alpha (A_0+\rho_0\sin\delta)A_{22}.
\label{eq:3_b}
\end{equation}
\end{subequations}
Here,
\begin{subequations}
\label{eq:3_eta}
\begin{eqnarray}
 \tilde{\eta}_0&=&(1-3\rho_0^2\sin^2\delta)\alpha -\gamma a_0,\\
 \tilde{\eta}_1&=&(1-3\rho_0^2\sin^2\delta)\alpha -\beta k^2
 -\gamma (a_0-a_1 k),\\
 \tilde{\eta}_2&=&(1-3\rho_0^2\sin^2\delta)\alpha -4\beta k^2
 -\gamma (a_0-2a_1 k).
\end{eqnarray} 
\end{subequations}
Equation~(\ref{eq:3_b}) suggests that there can be a traveling pattern
if both $A_0+\rho_0\sin\delta\ne 0$ and $A_{22}\ne 0$ are satisfied.

Now let us think about the SPs of
Eq.~(\ref{eq:3_0}) where  
$A_0+\rho_0\sin\delta = 0$ or $A_{22}=0$. For
the cases with $k\simeq 1$ and the parameter set used in
Fig.~\ref{fig:trvl}, we find
that there are no SPs with $A_0+\rho_0\sin\delta = 0$. 
Therefore, we concentrate on SPs with $A_{22}=0$, where
Eq.~(\ref{eq:3_0}) leads to
the following equations:
\begin{subequations}
\label{eq:3_1}
\begin{eqnarray}
 0&=&\tilde{\eta}_0A_0+C -\alpha\left[
A_0^2(A_0+3\rho_0\sin\delta) +\frac32 (A_0+\rho_0\sin\delta)
(A_1^2+A_{21}^2) -\frac34 A_1^2A_{21}
\right], \label{eq:3_1a}\\
 0&=&\tilde{\eta}_1A_1 -\alpha\left[
\frac34 A_1^3 +\frac32 A_1A_{21}^2
+3A_0A_1(A_0+2\rho_0\sin\delta) -3A_1A_{21}(A_0+\rho_0\sin\delta)
\right], \label{eq:3_1b}\\
 0&=&\tilde{\eta}_2A_{21} -\alpha\left[
\frac34 A_{21}^3 +\frac32 A_1^2A_{21}
+3A_0A_{21}(A_0+2\rho_0\sin\delta) 
-\frac32 A_1^2(A_0+\rho_0\sin\delta) 
\right]. \label{eq:3_1c}
\end{eqnarray} 
\end{subequations}
Here, we note that $A_1$ should not be zero. When
$A_1\ne 0$, Eq.~(\ref{eq:3_1b}) leads to
\begin{equation}
 A_1^2=\frac{4\tilde{\eta}_1}{3\alpha}-2\left[
A_{21}^2 +2A_0(A_0+2\rho_0\sin\delta) -2A_{21}(A_0+\rho_0\sin\delta)
\right].
\label{eq:3_2}
\end{equation}
Substituting Eq.~(\ref{eq:3_2}) into Eqs.~(\ref{eq:3_1a}) and
(\ref{eq:3_1c}), we obtain a pair of nonlinear simultaneous equations for
$A_0$ and $A_{21}$, which can be solved numerically.

At those SPs, the Jacobian of Eq.~(\ref{eq:3_0}) is
\begin{equation}
 J=\left(
\begin{array}{cccc}
 J_{11} & J_{12} & J_{13} & 0 \\
 2J_{12} & J_{22} & J_{23} & 0 \\
 2J_{13} & J_{23} & J_{33} & 0 \\
 0 & 0 & 0 & J_{44} 
\end{array}
\right),
\label{eq:Jacobi3}
\end{equation}
with the elements given by
\begin{subequations}
\label{eq:3_3}
\begin{eqnarray}
 J_{11}&=&\tilde{\eta}_0 -\alpha\left[ 3A_0^2 +6A_0\rho_0\sin\delta
+\frac32 (A_1^2+A_{21}^2) \right],\\
 J_{12}&=&-3\alpha A_1\left(
A_0+\rho_0\sin\delta -\frac12 A_{21} \right),\\ 
 J_{13}&=&-3\alpha \left[ (A_0+\rho_0\sin\delta)A_{21} 
-\frac14 A_1^2 \right],\\
 J_{22}&=&\tilde{\eta}_1 -\alpha\left[ \frac94 A_1^2 +\frac32 A_{21}^2
+3A_0(A_0+2\rho_0\sin\delta) -3A_{21}(A_0+\rho_0\sin\delta) \right],\\
 J_{23}&=&3\alpha A_1(A_0+\rho_0\sin\delta -A_{21}),\\
 J_{33}&=&\tilde{\eta}_2 -\alpha\left[ \frac94 A_{21}^2 +\frac32 A_1^2 
+3A_0(A_0+2\rho_0\sin\delta)\right],\\
J_{44}&=&\tilde{\eta}_2 -\alpha\left[ \frac34 (2A_1^2+A_{21}^2)
+3A_0(A_0+2\rho_0\sin\delta) +6A_{21}(A_0+\rho_0\sin\delta) \right].
\end{eqnarray} 
\end{subequations}
Equation~(\ref{eq:Jacobi3}) is a block-diagonal matrix. We can evaluate the
real parts of the 
eigenvalues, $\Lambda_1$, $\Lambda_2$, $\Lambda_3$, for the upper-left
$3\times 3$ matrix as well as $\Lambda_4=J_{44}$.
\begin{figure}
\includegraphics[width=8cm]{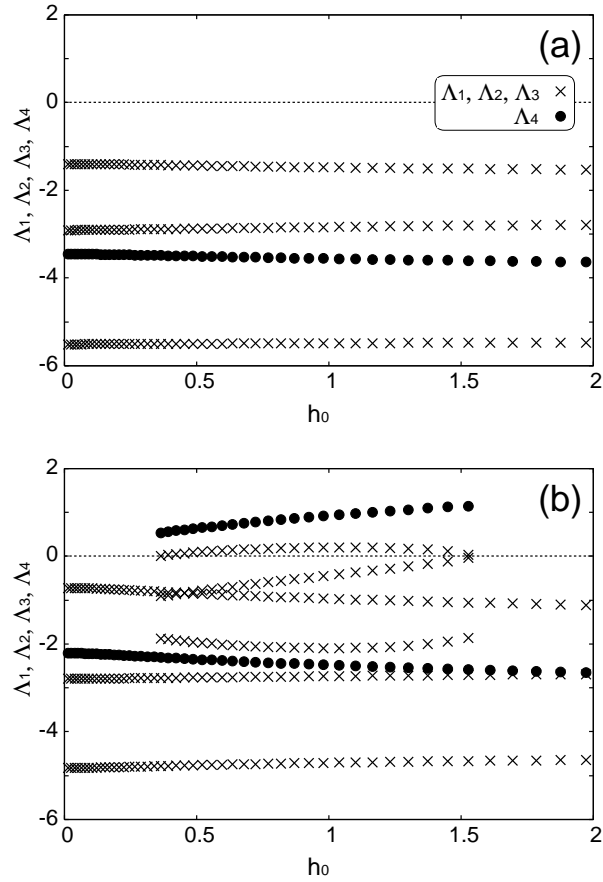}
\caption{\label{fig:eigen} The dependence of the real parts of
 the eigenvalues of Eq.~(\ref{eq:Jacobi3}) on the field
amplitude $h_0$: (a) $k=1.0$ and (b) $k=0.83$.} 
\end{figure}
The dependence of the real parts of the eigenvalues on the field
amplitude $h_0$ is shown 
in Fig.~\ref{fig:eigen}. Here, we take the values of the parameters,
$\alpha$, $\beta$, and $\gamma$, used in Fig.~\ref{fig:trvl}.
For $k=1.0$, all the real parts of the
eigenvalues ($\Lambda_1, \ldots,\Lambda_4$) are always negative. 
In other words, the
SPs are stable and a traveling pattern cannot appear.
For $k=0.83$, however, extra SPs appear in the region between 
$h_0\simeq 0.35$ and $h_0\simeq 1.5$. In that region,  
$\Lambda_4$ and one of the other three 
($\Lambda_1, \Lambda_2, \Lambda_3$) 
are positive. Incidentally, it is confirmed that 
$A_0+\rho_0\sin\delta\ne 0$ in the region.
This result suggests that a traveling pattern
can appear in a certain region of the field when $k=0.83$. 

\begin{figure}
\includegraphics[width=8cm]{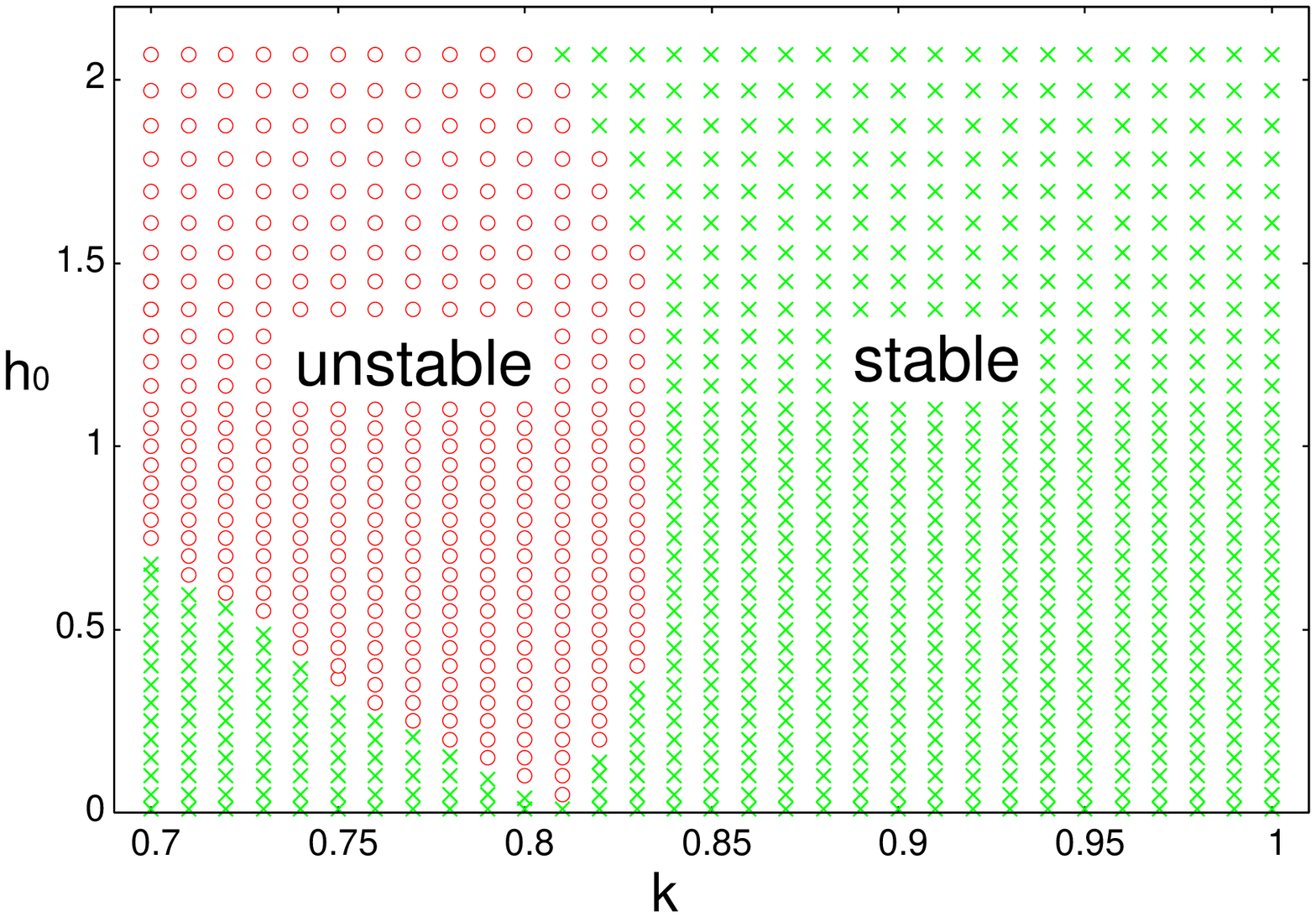}
\caption{\label{fig:k_h0} Stability diagram in $k$-$h_0$ space.
In the unstable area (red-circle points), a traveling pattern can appear.
In the stable area (green-cross points), the pattern cannot travel.} 
\end{figure}

Using the above analysis, we show a stability diagram in 
Fig.~\ref{fig:k_h0}. In the unstable area, where there is a branch with 
positive $\Lambda_4$, a traveling pattern can appear. In the stable area, 
where all the branches of $\Lambda_4$ have negative values, it cannot 
appear. The values of the parameters $\alpha$, $\beta$, and $\gamma$ are 
the same as ones in Fig.~\ref{fig:trvl}. The characteristic wave number 
$k_0$ depends on the ratio of $\beta$ and $\gamma$ (see Ref.~\cite{kudoS} 
for details), and $k_0=1$ in our case. Though it is expected that 
$k\simeq k_0$, the actual characteristic length in the simulations is 
larger than $2\pi /k_0$. In other words, $k<k_0$ in the actual numerical 
results, although a domain pattern with a small $k$ is not always 
realistic. 
Incidentally, the field larger than $h_0\simeq 1.5$ may be meaningless
since domain patterns should vanish under a strong field. 

\section{\label{sec:disc} Discussion}

The results in Fig.~\ref{fig:eigen} suggest that a traveling pattern can
exist for $k\le 0.83$ but not for $k=1.0$. This can be interpreted as
meaning that a traveling pattern should be a little fat.
In fact, the actual wave numbers of domain
patterns in our numerical simulations are a little less than $k=1$,
although $k=0.83$ seems too small.
We can say that this fact partly supports the theoretical results given
here.   

We have used very simple approximations, i.e. perfect parallel-stripe
structures without any distortion, to investigate the instabilities of
a traveling pattern. That may be one of the reasons why the present
analysis has suggested a traveling pattern with
$k$ smaller than that of the numerical results.
In our simulations, the traveling patterns do not have a perfect
parallel-stripe structure. If more complex and better approximations are
employed, the actual traveling patterns exhibited by numerical
simulations may be better explained.

In experiments, the perfect parallel-stripe structure is a realistic 
pattern. However, as mentioned above, the condition for a traveling 
pattern is tight even for such a simple structure. Traveling patterns 
with a more complex structure can be observed in experiments and the 
conditions of their appearance would be more complex than the present 
case. In any case, it is sure that a traveling pattern cannot appear 
without a rapid oscillating field.

\section{\label{sec:conc} Conclusions}

We have proposed two ways to describe 
magnetic domain patterns moving slowly under a rapidly
oscillating field. One gives a model in which rapidly oscillating terms
are averaged out. The time-averaged model can explain the existence of
the maximum values of the field where non-uniform domain patterns are
preferable. The other gives a model which includes a phase shift as the
delay of the response to the field. The phase-shifted model suggests the
existence of a traveling pattern which moves very slowly compared with
the time scale of the field. 
These two models have both merits and demerits. In other words, the
approximations to be employed depend on the phenomenon under
consideration.
We should choose a method suitable for the analysis of the phenomenon 
to be investigated.

Although we have focused on a traveling
pattern in this paper, these two methods are promising for applying to
many other domain patterns under a rapidly oscillating field.

\begin{acknowledgments}
The authors would like to thank M.~Mino for information about
 experiments and M.~I.~Tribelsky, M.~Ueda, and Y.~Kawaguchi for useful
 comments and discussion.    
 One of the authors (K.~K.) is supported by JSPS Research Fellowships
 for Young Scientists. 
\end{acknowledgments}

\appendix
\section{\label{sec:app} Numerical simulations}

The numerical procedures for time evolution are almost the same as
those of Refs.~\cite{kudoE,kudoS}. For time evolution, we use a
semi-implicit method: The exact solutions and the second order
Runge-Kutta method are used for the linear and nonlinear terms,
respectively. For a better spatial resolution, a pseudo-spectral method
is applied. In other words, we numerically calculate the time evolutions of
Eq.~(\ref{eq:A-C}) in Fourier space:
\begin{equation}
 \frac{\partial \phi_{\bm{k}}}{\partial t} =\alpha [\phi-\phi^3]_{\bm{k}}
-(\beta k^2 +\gamma G_{\bm{k}})\phi_{\bm{k}} + h(t)\delta_{\bm{k}},
\label{eq:a-1}
\end{equation} 
where $[\cdot]_{\bm{k}}$ denotes the convolution sum and  $G_{\bm{k}}$ is the
Fourier transform of $G({\bm{r}},0)$. Since we defined  
$G(\bm{r},0)\equiv 1/|\bm{r}|^3$,
\begin{equation}
 G_{\bm{k}}=a_0-a_1 k,
\end{equation}
where $k=|\bm{k}|$ and
\begin{equation}
 a_0=2\pi\int_d^{\infty} \frac{{\rm d}r}{r^2}, \quad a_1=2\pi.
\end{equation}
In the simulations, we set $d=\pi/2$, which results in $a_0=4$.

In Fig.~\ref{fig:trvl}, the parameters are given as $\alpha=2.0$,
$\beta=2.0$, and $\gamma=2\beta/a_1=2/\pi$. 
The frequency and amplitude of the field are
$\omega=2\pi\times 5\times 10^{-2}$ and $h_0=0.8$, respectively. 
The simulations are performed on a $128\times 128$ lattice
with periodic boundary conditions.
The snapshots in Fig.~\ref{fig:trvl} are the domain patterns at (a)
$5\times 10^3T$, (b)  
$10\times 10^3T$, and (c) $15\times 10^3T$, where $T=2\pi/\omega$.
If the amplitude is larger (for example, $h_0=0.9$, $0.95$, etc.), 
we can see 
a traveling pattern with a different structure  moves to a
different direction.

\end{document}